\documentstyle[12pt,amsfonts]{article}

\def\half{\textstyle{\frac{1}{2}}}
\def\quarter{\textstyle{\frac{1}{4}}}

\def\threebytwo{\textstyle{\frac{3}{2}}}
\def\H{{\cal H}}
\def\ol{\overline}

\def\Tr{{\rm Tr}}

\def\tint{{\textstyle\int}}

\def\s{\hskip.08em}

\def\b{\begin{eqnarray}}  
\def\e{\end{eqnarray}}    
\def\bn{\begin{eqnarray}}  
\def\<{\langle}
\def\>{\rangle}
\def\ss{\sigma}

\def\no{\nonumber}

\def\{{\lbrace}
\def\}{\rbrace}
\begin{document}
\title{Extension of Berezin-Lieb Inequalities}
\author{John R.~Klauder
\footnote{Electronic mail: klauder@phys.ufl.edu}\\
Department of Physics and \\ Department of Mathematics\\
University of Florida\\
Gainesville, FL 32611\\
and\\
Bo-Sture K.~Skagerstam\footnote{Electronic mail: bo-sture.skagerstam@ntnu.no}\\
 Department of
Physics\\ The Norwegian University of Science and Technology\\
N-7491 Trondheim, Norway}
\date{}    
\maketitle
\begin{abstract}
The Berezin-Lieb inequalities provide upper and lower bounds for a partition function based on phase space integrals that involve the Glauber-Sudarshan  and Husimi representations, respectively. Generalizations of
these representations have recently been introduced by the present authors, and in this article, we extend the
use of these new representations to develop numerous analogs of the Berezin-Lieb inequalities that  may offer improved bounds. Several
examples illustrate the use of the new inequalities. Although motivated  by problems in quantum mechanics, these results may also find applications in time-frequency analysis, a valuable cross fertilization that has been profitably used at various times in the past.
\end{abstract}
\section{Introduction}

The Berezin-Lieb inequalities offer upper and lower bounds for
partition functions of elementary quantum systems. In particular,
for a system composed of a single canonical degree of freedom, let
$P$ and $Q$ denote canonical Heisenberg variables, fulfilling the
commutation relation $[Q,P]=iI$, in units where $\hbar=1$. Let
$|\s0\>$ denote the normalized ground state of an elementary oscillator for
which $(Q+iP)|\s0\>=0$. Canonical coherent states for this system are
taken to be states of the form (see, e.g., Refs. \cite{Klauder_1985})
  \b |p,q\>\equiv U[p,q]\s|\s0\>\;,\hskip1cm U[p,q]\equiv e^{i(p\s Q-qP)} \e
for all $(p,q)\in{\mathbb R}^2$, where $U[p,q]$ denotes the unitary
Weyl operator. Let $\H=\H(P,Q)$ denote the Hamiltonian for the
system in question. The corresponding classical Hamiltonian is denoted by $H_{cl}(p,q)$. We introduce two well-known symbols associated
with $\H$, namely, the Husimi \cite{H} symbol $H_H(p,q)$ defined by
    \b  H_H(p,q)\equiv
    \<p,q|\s\H(P,Q)\s|p,q\>=\<0|\s\H(P+p,Q+q)\s|\s0\>\;; \label{H} \e
    and the Glauber-Sudarshan \cite{G-S} symbol $H_{G-S}(p,q)$ implicitly
    defined by the operator representation
     \b \H(P,Q)=\int H_{G-S}(p,q)\,|p,q\>\s\<p,q|\,dp\s dq/2\pi\;.
     \e
It follows from Eq.~(\ref{H}) that these two symbols are related by the
integral equation
   \b &&H_H(p',q')=\int |\<p',q'|p,q\>|^2\,H_{G-S}(p,q)\,dp\s
   dq/2\pi\no\\
   &&\hskip1.85cm=\int
   e^{-[(p'-p)^2+(q'-q)^2]/2}\,H_{G-S}(p,q)\,dp\s dq/2\pi\;. \e

 Armed with these definitions, the Berezin-Lieb inequalities
\cite{ber,lie} read
 \b \int e^{-\beta\s H_H(p,q)}\,dp\s dq/2\pi\le
 {\rm Tr}[e^{-\beta\s\H(P,Q)}]\le \int e^{-\beta\s
 H_{G-S}(p,q)}\,dp\s dq/2\pi \;. \label{bl}\e
 In what follows we will implicitly rederive this inequality
 as a special example of our generalizations.

 The purpose of the present paper is to extend such inequalities by
 offering infinitely many additional symbol pairs that can stand
 in place of the Husimi and Glauber-Sudarshan symbols in Eq.~(\ref{bl}), thereby
 generalizing the original Berezin-Lieb inequalities.

 \section{Multiple phase-space symbols}
 In a recent paper \cite{klsk}, the authors have introduced a wide class
 of phase-space symbols that are analogues of the Husimi and
 Glauber-Sudarshan dual pair. Let us first recall the principal
 elements of that study specialized to the discussion at hand.

 We first introduce a nonnegative, trace-class operator $\sigma=\sigma^\dagger\ge0$
 which we normalize so that $\Tr(\sigma)=1$.
      Such operators have the generic form given by
        \b \sigma=\sum_{l=1}^\infty c_l\,|\s b_l\>\s\<b_l|\;, \label{c} \e
        where $\{|\s b_l\>\}_{l=1}^\infty$
        denotes a complete orthonormal sets of vectors, and the
        coefficients $\{c_l\}_{l=1}^\infty$ satisfy the conditions
        $c_l\ge0$ and $\Sigma_{l=1}^\infty c_l=1$. In short, $\sigma$
        enjoys all the properties to be a density matrix.

       We shall make use of the function $\Tr(U[k,x]\s\ss)$ defined for all
       $(k,x)$ in phase space, and we restrict $\ss$ so that the expression
          \b
\label{eq_restriction}
\Tr(U[k,x]\,\ss)\not=0 \e
          for all $(k,x)\in{\mathbb R}^2$.

          We next recall the Weyl representation of operators given by
          \b A=\int {\tilde A}(k,x)\s U[k,x]\,dk\s dx/2\pi \;, \e
          where
          \b
\label{eq:A_tilde}
{\tilde A}(k,x)\equiv\Tr(U[k,x]^\dagger\s A)\;.  \e
          Given two such operators $A$ and $B$, it follows that
   \b    \Tr(A^\dagger\s B)=\int {\tilde A}(k,x)^*\s{\tilde B}(k,x)\,dk\s dx/2\pi \;.\label{W} \e
           In terms of the double Fourier transformation, given by
           \b A(p,q)=\int e^{i(qk-px)}\,{\tilde A}(k,x)\,dk\s
           dx/2\pi\;,\e
           and likewise for $B(p,q)$, it also follows that
           \b \Tr(A^\dagger\s B)=\int A(p,q)^*\s B(p,q)\,dp\s
           dq/2\pi \;. \e

            We next modify the symmetric expression for
            $\Tr(A^\dag\s B)$ given by Eq.~(\ref{W})  so that
            \b
\label{eq_dividebyzero} &&\hskip-.6cm\Tr(A^\dag\s B)= \int \{\s\frac{{\tilde A}(k,x)^*}
            {\Tr(U[k,x]\s\ss)}\s\} \,\{\Tr(U[k,x]\s\ss){\tilde
            B}(k,x)\}\,dk\s dx/2\pi\nonumber\\
          &&\hskip1.05cm =\int \{\frac{{\tilde A}(k,x)}
          {\Tr(U[k,x]^\dag\s\ss)}\s\}^*\,
     \{\s\Tr(U[k,x]\s\ss){\tilde
            B}(k,x)\s\}\,dk\s dx/2\pi\no\\
             &&\hskip1.05cm\equiv \int {\tilde A}_{-\ss}(k,x)^*\,
             {\tilde B}_\ss(k,x)\,dk\s dx/2\pi\nonumber\\
        &&\hskip1.05cm\equiv \int A_{-\ss}(p,q)^*\,B_\ss(p,q)\,dp\s
          dq/2\pi\;.\e
    In the final line we have introduced the Fourier
    transform of the symbols in the line above. We next show that
    there are
    alternative expressions involving the symbols $A_{-\ss}(p,q)$
    and $B_\ss(p,q)$ directly in their own space of definition
    rather than implicitly through a Fourier transformation.

    We begin first with the symbol $B_\ss(p,q)$. In particular,  we
    note that
     \b &&\hskip-.75cm B_\ss(p,q)=\int e^{i(kq-xp)}\,\Tr(U[k,x]\s\ss)\,{\tilde
            B}(k,x)\,dk\s dx/2\pi\nonumber\\
       &&\hskip.8cm
= \int\Tr(U[p,q]^\dag\,U[k,x]\,U[p,q]\s\ss)
       \,\Tr(U[k,x]^\dag\s B)\,dk\s dx/2\pi\nonumber\\
       &&\hskip.8cm
= \int \Tr(U[k,x]\,U[p,q]\s\ss\,U[p,q]^\dag)
       \,\Tr(U[k,x]^\dag\s B)\,dk\s dx/2\pi\nonumber\\
       &&\hskip.8cm=\Tr(U[p,q]\s\ss\,U[p,q]^\dag\s B)\;, \e
       where in the second line we have used the Weyl form of the commutation
       relations, and in the last line we have used the Weyl representation Eq.~(\ref{W}), which
       leads us to the desired expression for $B_\ss(p,q)$. This
       expression is the sought for
       generalization of the Husimi representation; indeed, if $\ss=|\s 0\>\s\<0|$
       it follows immediately that
         \b B_{\ss}(p,q)\hskip-1.3em&&=\Tr(U[p,q]\s|\s 0\>\s\<0|\s
         U[p,q]^\dag\s B)\nonumber\\ &&=\<p,q|\s B\s|p,q\>=B_H(p,q)\;.\e

     For general $\ss$, to find the expression for $A_{-\ss}(p,q)$ we appeal to the
     relation
        \b &&\Tr(A^\dag\s B)=\int A_{-\ss}(p,q)^*\,B_\ss(p,q)\,dp\s dq/2\pi\nonumber \\
       &&\hskip1.7cm
=\int A_{-\ss}(p,q)^*\Tr(U[p,q]\s\ss\,U[p,q)]^\dag\s B)\,dp\s
dq/2\pi\;, \e
        an equation, which, thanks to its validity for all suitable operators $B$,
         carries the important implication that
\b A\equiv \int A_{-\ss}(p,q)\,U[p,q]\s\ss\,U[p,q]^\dag\,dp\s
dq/2\pi\;.\label{u10} \e
        Observe that this equation implies a very general operator
        representation as a linear superposition of basic operators
        given by
            $U[p,q]\s\ss\,U[p,q]^\dag $,
            for a general choice of $\ss$.

        Equation (\ref{u10}) for $A$ is the sought for generalization of the
        Glauber-Sudarshan representation; indeed, if $\ss=|\s
        0\>\s\<0|$,
        it follows immediately that
          \b &&A=\int A_{-\ss}(p,q)\,U[p,q]\s|\s 0\>\s\<0|\s U[p,q]^\dag\,dp\s dq/2\pi\nonumber\\
       &&\hskip.38cm
=\int A_{-\ss}(p,q)\;|p,q\>\s\<p,q|\,dp\s dq/2\pi \no \\
        &&\hskip.38cm
=\int A_{G-S}(p,q)\;|p,q\>\s\<p,q|\,dp\s dq/2\pi\;. \e
   Once again there is a direct connection between the
   generalization of the
   Husimi representation, $A_\ss(p,q)$, and the generalization of the
    Glauber-Sudar- shan representation, $A_{-\ss}(p,q)$. In
    particular, it follows that
    \b A_\ss(r,s)\hskip-1.3em&&=
    \int A_{-\ss}(p,q)\s\Tr(U[r,s]\s\ss U[r,s]^\dag U[p,q]\s\ss
    U[p,q]^\dag)\s dp\s dq/2\pi\no\\
      &&\hskip-1.5em=\int A_{-\ss}(p,q)\s
      \Tr(U[r-p,q-s]\s\ss U[r-p,q-s]^\dag\ss)
      \s dp\s dq/2\pi. \label{q9}\e
      This equation is a convolution, which just reflects the
      multiplicative connection between these two symbols in
      Fourier space.

      \section{Derivation of inequalities}
      Let $\{|\s r\>\}_{r=1}^\infty$ denote an arbitrary, complete, orthonormal
      basis. Consider the expression [cf., Eq.~(\ref{c})]
        \b f(p,q|\s r)\hskip-1.3em&&\equiv \<r|U[p,q]\s\ss\,U[p,q)]^\dag |\s r\>\no\\
        &&=\sum_{l=1}^\infty\,c_l\,|\<r|\s U[p,q]\s|\s b_l\>|^2\;.
        \e
        It follows that
        \b \int f(p,q|\s r)\,dp\s dq/2\pi=1\;, \e
        and also that
        \b \sum_{r=1}^\infty f(p,q|\s r)=1\;. \e
        We can interpret these results in two different ways:
        On the
        one hand, $f(p,q|\s r)$ is a {\it probability density} on ${\mathbb
        R}^2$ for each value of $r$; on the other hand, $f(p,q|\s r)$
        forms a {\it discrete probability} on $\{1,2,3,\ldots\}$ for each
        phase-space point $(p,q)$.

\subsection{Jensen's inequality}
        The Jensen inequality \cite{jen} applies to convex functions $\phi(\s x\s)$---such as
        $e^{-\beta x}$---and arbitrary probability distributions on
        $x\in{\mathbb R}$. If $\<(\s\cdot\s)\>$ denotes an average
        over that probability distribution, then the Jensen
        inequality reads
          \b \phi(\<\s x\s\>)\le \<\phi(\s x\s)\>\;,  \e
          or, in particular,
          \b  e^{-\beta\<\s x\s\>}\le\<\s e^{-\beta x}\s\>\;. \e
          This equation will be important in what follows.

        Let $\H$ denote the Hamiltonian with a discrete spectrum $\{\mu_r\}_{r=1}^\infty$
        and an associated set of
        eigenvectors $\{|\s r\>\}_{r=1}^\infty$
         such that
          \b \H\s|\s r\>=\mu_r\s|\s r\>\;. \e
          It also follows that
          \b \H=\sum_{r=1}^\infty \mu_r\,|\s r\>\s\<r|\;. \e

        Following Lieb \cite{lie},  we first observe that
        \b &&\hskip-1cm\<r|\s e^{-\beta\s\H}\s|\s r\> =\exp[-\beta\<r|\H|\s r\>]\no\\
        &&\hskip1.1cm=\exp[\s-\beta\tint H_{-\ss}(p,q)\,f(p,q|\s r)\,dp\s dq/2\pi\,]\no\\
        &&\hskip1.1cm\le\int e^{-\beta\s H_{-\ss}(p,q)}\,f(p,q|\s r)\,dp\s
        dq/2\pi\;. \e
        Summing on $r$ leads to
         \b \Tr( e^{-\beta\s\H}\s)\le\int e^{-\beta\s H_{-\ss}(p,q)}\,dp\s
        dq/2\pi\;. \e
        Second, we learn that
         \b &&\exp[-\beta\s
         H_\ss(p,q)]=\exp[-\beta\Sigma_r\s\mu_r\,f(p,q|\s r)\s]\no\\
         &&\hskip3.05cm\le\s\sum_{r=1}^\infty
         \exp[-\beta\s\mu_r]\,f(p,q|\s r)\;. \e
         Integrating over ${\mathbb R}^2$ leads to
         \b \int e^{-\beta\s H_\ss(p,q)}\,dp\s dq/2\pi\le\Tr( e^{-\beta\s\H}\s)\;. \e

         Above we have two separate inequalities, one an upper
         bound, the other a lower bound. These bounds apply for {\it
         any} choice of $\ss$ that fits our requirements, and so we can decouple the choice of
         $\ss$ and assert that $\ss$ can be chosen independently in the
         two cases. In summary, therefore, we have established the
         inequalities
           \b   \label{k_bounds} \int e^{-\beta\s H_{\ss'}(p,q)}\,dp\s dq/2\pi\le
           \Tr( e^{-\beta\s\H}\s)\le\int e^{-\beta\s H_{-\ss}(p,q)}\,dp\s
        dq/2\pi\;,   \e
        where $\ss'$ and $\ss$ may be chosen independently of each
        other. This possibility permits optimizing both bounds by
        taking the supremum over the lower bound and taking the
        infimum over the upper bound. The bounds as given by Eq.~(\ref{k_bounds}) now lead to upper and lower bounds, respectively, of the free energy $F(\beta) \equiv -\ln Z(\beta)/\beta$, where $Z(\beta)$ denotes the
         partition function, as well as bounds on the ground-state energy $E_0$ since $E_0 = \lim_{\beta \rightarrow \infty }  F(\beta)$.
\section{Symbols for the lower bound}
   We focus on the symbol
  \b H_\ss(p,q)\hskip-1.3em&&=\Tr(U[p,q]\s\ss\s U[p,q]^\dagger\s\H(P,Q)\s) \nonumber\\
  &&=\Tr(\H(P+p,Q+q)\s\ss\s)\;. \e
  For simplicity, we introduce the shorthand notation that
    \b \ol{(\s\cdot\s)}\equiv \Tr(\s(\s\cdot\s)\s\ss\s)\;. \e
    In that case we find, e.g., that
        \b (q)_\ss \equiv\Tr(\s( Q+q)\s\ss\s)\equiv q+\ol{Q} \;,\e
        where the notation $(q)_\ss$ is the symbol
        $H_\ss(p,q)$ when the operator $\H$ is simply $Q$.
        Below we list a table of symbols needed for our present
        purposes:
        \b&& (q)_\ss=q+\ol{Q} \; , \no\\
&& (p)_\ss=p+\ol{P} \; ,\no\\
&& (q^2)_\ss=q^2+2q\ol{Q}+\ol{Q^2}\; ,\no\\
&& (p^2)_\ss=p^2+2p\ol{P}+\ol{P^2} \; ,\no\\
&& (qp)_\ss=qp+q\ol{P}+p\ol{Q}+\overline{QP} \; ,\no\\
&& (pq)_\ss=pq+p\ol{Q}+q\ol{P}+\ol{PQ} \; ,\no\\
&& (q^4)_\ss=q^4+4q^3\ol{Q}+6q^2\ol{Q^2}+4q\ol{Q^3}+\ol{Q^4}\; ,\no\\
&& (p^4)_\ss=p^4+4p^3\ol{P}+6p^2\ol{P^2}+4p\ol{P^3}+\ol{P^4}\; ,\no\\
&&(q^2p^2)_\ss=q^2p^2+2pq^2\ol{P}+2qp^2\ol{Q}+q^2\ol{P^2}+p^2\ol{Q^2}+4qp\ol{QP}\no\\
&&\hskip2cm +2q\ol{QP^2} +2p\ol{Q^2P}+\ol{Q^2P^2}\; ,\no\\
&&(p^2q^2)_\ss=p^2q^2+2pq^2\ol{P}+2qp^2\ol{Q}+q^2\ol{P^2}+p^2\ol{Q^2}+4pq\ol{PQ}\no\\
&&\hskip2cm +2q\ol{P^2Q} +2p\ol{PQ^2}+\ol{P^2Q^2}\; . \e
Note that on the left-hand side the order matters, i.e.,
$(qp)_\ss\not=(p\s q)_\ss$, etc. We also notice that for the quadratic symbols
\b
\label{lower_quad}
&& (q^2)_\ss=(q+\ol{Q})^2+ \Delta(Q)\; ,\no\\
&& (p^2)_\ss=(p+\ol{P})^2+\Delta(P)\; ,\no \no\\
&& \half[\s(qp)_\ss+(pq)_\ss\s]=(q+\ol{Q})(p+\ol{P})+\Delta(Q ,P)\; ,
\e
in terms of the variances $\Delta({\cal O})\equiv\ol{{\cal O}^2}-\ol{{\cal O}}^2$ and $\Delta({\cal O}_1 ,{\cal O}_2)\equiv(\ol{{\cal O}_1{\cal O}_2+{\cal O}_2{\cal O}_1})/2-\ol{{\cal O}_1}\,\s\ol{{\cal O}_2}$.
For a conventional minimal uncertainty state, e.g.,
$\Delta(Q)\Delta(P)=1/4$ and $\Delta(Q,P)=0$.

We also introduce a special-case table based on a symmetry we shall
impose on $\ss$, and to be made use of below, namely, that all odd-order averages vanish, i.e.,
$\ol{Q}=\ol{Q^3}=\ol{P}=\ol{P^3}=\ol{Q^2P}=0$,  etc. This
special-case table reads
       \b \label{special}&& (q)_\ss=q \; ,\no\\
&& (p)_\ss=p \; ,\no\\
&& (q^2)_\ss=q^2+\ol{Q^2} \; ,\no\\
&& (p^2)_\ss=p^2+\ol{P^2} \; ,\no\\
&& (qp)_\ss=qp+\overline{QP} \; ,\no\\
&& (pq)_\ss=pq+\ol{PQ} \; ,\no\\
&& (q^4)_\ss=q^4+6q^2\ol{Q^2}+\ol{Q^4}\; ,\no\\
&& (p^4)_\ss=p^4+6p^2\ol{P^2}+\ol{P^4}\; ,\no\\
&&(q^2p^2)_\ss=q^2p^2+q^2\ol{P^2}+p^2\ol{Q^2}+4qp\ol{QP}+\ol{Q^2P^2}\; ,\no\\
&&(p^2q^2)_\ss=p^2q^2+q^2\ol{P^2}+p^2\ol{Q^2}+4pq\ol{PQ}+\ol{P^2Q^2}\; .\e

\section{Symbols for the upper bound}
The construction of the upper limit is somewhat more involved than that for the lower limit.
We start with Eq.~(\ref{u10}), which is
   \b A=\int A_{-\ss}(p,q)\,U[p,q]\s\ss U[p,q]^\dagger\,dp\s dq/2\pi\;. \e
  For reasons of clarity  we limit ourselves to a a number-operator diagonal form for $\ss$, i.e.
   $\ss=\Sigma_{n=0}^\infty c_n\s|n\>\<n|$, $c_n\ge0$, and $\Sigma_{n=0}^\infty\s c_n=1$, where $N\s|n\>=n\s|n\>$ for the number
   eigenstates $\{\s|n\>\s\}$. We learn that in general
      \b A \hskip-1.3em
 &&=\Sigma_{n=0}^\infty \s c_n\s\int A_{-\ss}(p,q)\,U[p,q]|n\>\<n| U[p,q]^\dagger\,dp\s dq/2\pi\no\\
        && \equiv\Sigma_{n=0}^\infty \s c_n\s\int A_{-\ss}(p,q)\,|p,q;n\>\< p,q;n|\,dp\s dq/2\pi\;, \e
in terms of the so called semi-coherent states or displaced coherent states $|p,q;n\> \equiv U[p,q]|n\>$ (see, e.g., \cite{Carruthers_1965}).
         To see what this means, let us take a simple example  with $A=P^2+Q^2$. Since an operator is determined by its expectation value in canonical coherent states, it is sufficient to consider the Husimi symbol $A_H(p,q)$ as given by Eq.~(\ref{H}), i.e.,

         \b \<r,s;0|\s(P^2+Q^2)\s|r,s;0\> 
=\<0|\s[\s (P+r)^2+(Q+s)^2\s]\s|0\>=(r^2+s^2)+1\no\\
                 \equiv\Sigma_{n=0}^\infty \s c_n\s\int [(p+r)^2+(q+s)^2+k_2]\,p_n(p,q)\,dp\s dq/2\pi \; ,~~~~~~~
        \e
   where
\b
\label{eq:p_n}
p_n(p,q) \equiv |\<0|p,q;n\>|^2= e^{-(p^2+q^2)/2}(p^2+q^2)^n/2^nn! \; .
\e
       Here, we have made use of the Ansatz
\b
\label{HQQ}
(P^2+Q^2)_{-\ss}(p,q)=p^2+q^2+k_2 \; ,
\e
where $k_2$ is a constant to be determined, and  we immediately learn that
            \b
\label{eq:k2}
k_2\equiv -1-2\s\Sigma_{n=0}^\infty c_n \s n \equiv -1 - 2\s \ol{n}\;, \label{e42} \e
where we have defined mean-values $\ol{f(n)}\equiv \Sigma_{n=0}^\infty c_n f(n)$.  It now, e.g.,
follows that the right-hand side of Eq.~(\ref{u10}), with the upper symbol as given by Eqs.~(\ref{HQQ}) and (\ref{e42}),
has $|n\>$ as an eigenstate with eigenvalue $2\s n+1$. It is not entirely trivial to verify this explicitly, but it follows using the properties of displaced coherent states as well as properties of the conventional associated Laguerre polynomials $L_n^m$:
\b
\label{Laguerre}
L_n^m(x) = \sum_{k=0}^n (-1)^k \frac{(n+m)!x^k}{(n-k)!k!(m+k)!} \; .
\e

           In like fashion, it follows for $A=(P^2+Q^2)^2$ and the corresponding Husimi symbol that
           \b \<r,s;0|\s(P^2+Q^2)^2\s|r,s;0\>
=\<0|[(P+r)^2+(Q+s)^2]^2|0\> =(r^2+s^2+1)^2\no\\
         \equiv\Sigma_{n=0}^\infty \s c_n\s\int [((p+r)^2+(q+s)^2)^2+k_4((p+r)^2+(q+s)^2)+k_6] \no \\
\times\s p_n(p,q)\,dp\s dq/2\pi \;,\hskip6.3em \e
expressed in terms of the (assumed) symbol
\b
\label{H_squared}
((P^2+Q^2)^2)_{-\ss}(p,q)=(p^2+q^2)^2 + k_4(p^2+q^2)+k_6 \; .
\e
One now finds, making use of Eq.~(\ref{eq:p_n}), that
       \b
\label{eq:k4}
k_4=2-8\s\Sigma_{n=0}^\infty c_n\s(n+1)= -6 -8\s\ol{n} \; , \e
and
\b
\label{eq:k6}
k_6\hskip-1.3em&& =1-4\Sigma_{n=0}^\infty\s c_n\s (n+1)(n+2)-2k_4\Sigma_{n=0}^\infty\s c_n\s (n+1)\no \\
    &&=5+16\ol{n}+16\ol{n}^2-4\ol{n^2}\;.\hskip6.3em
\e

In a similar manner and for $A=Q^4$, we can write
\b
\label{Q4}
(Q^4)_{-\ss}(p,q)=q^4 + a_2\s q^2+a_4 \; ,
\e
where
       \b a_2= -3(1+2\s\ol{n}) \; , \e
and
          \b a_4=3\s(\s\s\quarter+\threebytwo\ol{n}+2\s\ol{n}^2-\half\ol{n^2}\s\s) \; . \e

                 The expressions above now relate the standard symbols to
                the generalized symbols. Extension of these expressions to other polynomials in $P$ and $Q$ is straightforward.

  \section{Examples}

With the special choice  for $\ss$ considered above, i.e.,
   $\ss=\Sigma_{n=0}^\infty c_n\s|n\>\<n|$, $c_n\ge0$, and $\Sigma_{n=0}^\infty\s c_n=1$, we will now consider some specific examples in order to illustrate the use of the generalized upper and lower symbols. We first remark that in the trivial case of an harmonic oscillator with $H=(P^2+Q^2)/2$, such that $Z(\beta)= 1/[2\sinh(\beta/2)]$,  the lower symbol Eq.~(\ref{lower_quad}) and the upper symbol Eq.~(\ref{HQQ}), together with the bounds Eq.~(\ref{k_bounds}), lead to the expression  
\b e^{-\beta(\Delta(P)+\Delta(Q))/2}/\beta  \leq Z(\beta)\leq e^{\beta(1/2+\ol{n})/2}/\beta, \e
which, obviously, is true.  We can optimize this expression
in the form
  \b
\label{h_bounds}
e^{-\beta /2}/\beta \leq Z(\beta) \leq e^{\beta/2}/\beta\;. \e
  From the corresponding lower bound we then obtain an upper bound on the ground-state energy $E_0\leq 1/2$ since
$E_0 = -\lim_{\beta \rightarrow \infty }  \ln Z(\beta)/\beta$.
In the high-temperature limit, i.e., $\beta \rightarrow 0 $, the bounds in Eq.~(\ref{h_bounds}) exactly reproduce the classical Gibbs partition function $Z_{cl}(\beta)/2\pi =1/\beta$ taking the  fundamental
phase-space volume $2\pi$ into account and making use of
\b
\label{Gibbs}
Z_{cl}(\beta)=\int e^{-\beta\s H_{cl}(p,q)}\,dp\s dq \; ,
\e
with, of course,  $H_{cl}(p,q)=(p^2+q^2)/2$.

\subsection{A non-linear oscillator}
Here we consider Hamiltonians of the form $\H=\H(N)$, where $N$ is the usual number operator. We study this example more for its
ease of analysis and pedagogical value.
We choose as our example $\H=(N-a)(N-b)$. Such a form of an Hamiltonian has its roots in, e.g., the description of a single-mode non-linear Kerr-medium in quantum optics or a single vibrational mode beyond the harmonic approximation. We make the choice $a=1$ and $b=5$. We observe that the partition function $Z(\beta)=\sum_{n=0}^{\infty}\exp[- \beta(n-1)(n-5)]$ then has the form $Z(\beta) \simeq \exp(4\beta)$ for large values of $\beta$. A straightforward application of Poisson re-summation techniques also leads to the behavior $Z(\beta) \simeq \sqrt{\pi /\beta}/2$ for small values of $\beta$, which corresponds to the high-temperature limit of the classical partition function $Z_{cl}/2\pi$ using Eq.~(\ref{Gibbs}) with $H_{cl}=(p^2+q^2)^2/4-7(p^2+q^2)/2+33/4$.

We may then combine these factors for $\H=(N-1)(N-5)$ at hand by noting that
  \b &&(N-1)(N-5)=\quarter\s (P^2+Q^2-1)^2-6\s\half\s(P^2+Q^2-1)+5\no\\
    &&\hskip1.5cm=\quarter(P^4+Q^4+P^2Q^2+Q^2P^2)-7\s\half(P^2+Q^2)+33/4\;.\e
    Consequently,
    \b H_\ss(p,q)\hskip-1.3em&&=
    \quarter[(p^4)_\ss+(q^4)_\ss+(p^2q^2)_\ss+(q^2p^2)_\ss]-7\s\half[(p^2)_\ss+(q^2)_\ss]+33/4\no\\
      &&=\quarter
      [\,p^4+6p^2\ol{P^2}+\ol{P^4}+q^4+6q^2\ol{Q^2}+\ol{Q^4}+q^2p^2+q^2\ol{P^2}+p^2\ol{Q^2}\no\\
      &&\hskip0.2cm +4qp\ol{QP}+\ol{Q^2P^2}+
p^2q^2+q^2\ol{P^2}+p^2\ol{Q^2}+4pq\ol{PQ}+\ol{P^2Q^2}\,]\no\\
 &&\hskip2cm -7\s\half[\,p^2+\ol{P^2}+q^2+\ol{Q^2}\,]+33/4\;.\e

Since we have restricted our choice of $\ss$  so that
 it is only a function of $N$, i.e., $\ss=\ss(N)$,  $\ss$  has now a
 symmetry that makes $\ol{Q^2}=\ol{P^2}\equiv C_2$,
 $\ol{P^4}=\ol{Q^4}\equiv C_4$, $\ol{Q^2P^2}=\ol{P^2Q^2}\equiv
 C_{22}$, and importantly that $\ol{QP}+\ol{PQ}=0$. The three constants $C_2,\,C_4,\,C_{22}$
 are the only remnant of $\ss$ in
 $H_\ss(p,q)$, and of necessity, they satisfy $C_2\ge 1/2$, $C_4\ge C_2^2$, and $C_4\ge C_{22}$.
With the restriction $\ss=\ss(N)$  we can actually be more precise and write
\b \label{Cdetails}
C_2= \half + {\bar n}\;, \;
    C_{22} =  \half(\ol{n^2} +{\bar n} +{\textstyle \frac{1}{2}})\;,\;
C_{4} =  {\textstyle \frac{3}{2}}({\ol{n^2} }+{\bar n} +{\textstyle \frac{1}{2}})    \;.
 \e
 Putting this information together, we find that
   \b H_\ss(p,q)=\quarter\s(p^2+q^2)^2+K_1\s(p^2+q^2)+K_2\;,  \e
   where
      \b &&K_1\equiv {\textstyle \frac{7}{4}}(C_2-2)= {\textstyle \frac{7}{4}}( {\bar n}-{\textstyle \frac{3}{2}})\;,   \no\\
      && K_2\equiv C_4 + \half C_{22} -7C_2+{\textstyle \frac{33}{4}} = \ol{(n-3)^2}-4     \;. \e

   We note the fact that $H_\ss(p,q)$ is a function only of the combination $(p^2+q^2)$
   on the basis of our restriction that $\ss=\ss(N)$. It follows,
   therefore, that the lower bound of interest is given by
     \b
\label{Kerr} \int \exp\{-\beta\s H_\ss(p,q)\}\,dp\s
     dq/2\pi=\half\int_0^\infty \exp\{-\beta[\quarter s^2+K_1
     \s s +K_2\s]\} ds\;, \e
     where we have passed to polar coordinates and set
     $s\equiv(p^2+q^2)$. The upper  
     bound integral is a function of
     $\beta$ as well as the $\ss$-parameters, $C_2,\,C_4$, and $C_{22}$, i.e., the independent  mean-value ${\bar n}$ and dispersion $\ol{(n- {\bar n})^2}$ parameters.

 The lower bound of Eq.~(\ref{k_bounds}) together with Eq.~(\ref{Kerr}) now leads to the lower bound $\sqrt{\pi/\beta}/2\leq Z(\beta)$ as $\beta \rightarrow 0$.  This lower bound again corresponds to the high-temperature limit for the classical partition function $Z_{cl}(\beta)/2\pi$. By making use of
$E_0 = -\lim_{\beta \rightarrow \infty }  \ln Z(\beta)/\beta$, Eq.~(\ref{Kerr}) leads to the upper limit
$E_0 \leq -4$ using the state $\ss=|3\>\<3|$. We observe that such a state will not strictly satisfy the restriction imposed by Eq.~(\ref{eq_restriction}) since  $\Tr(U[k,x]\,\ss)$ then will be zero at isolated points away from the origin $k=x=0$. But, in fact, the restriction Eq.~(\ref{eq_restriction}) is then not required if $A$ is a polynomial in $P$ and  $Q$ since the symbol ${\tilde A}(k,x)$ as defined in Eq.~(\ref{eq:A_tilde}) will involve derivatives of delta-functions with support at the origin  \cite{Klauder_1964,klsk}.

The upper bound of Eq.~(\ref{k_bounds}), using Eqs.~(\ref{HQQ}) and (\ref{H_squared}), now leads to
\b
\label{Kerr_upper}
Z(\beta) \leq \half\int_0^\infty \exp\{-\beta[\quarter (s^2+k_4s+k_6)- {\textstyle{\frac{7}{2}}}(s+k_2)+
 {\textstyle{\frac{33}{4}}}    ]\} ds\;,
\e
where the parameters $k_2$, $k_4$ and $k_6$ are given by the equations (\ref{eq:k2}), (\ref{eq:k4}) and (\ref{eq:k6}), respectively. It is now evident again that Eq.~(\ref{Kerr_upper}) reproduces the high-temperature limit of the classical partition function $Z_{cl}(\beta)\simeq \sqrt{\pi /\beta}/2$. The upper bound of Eq.~(\ref{k_bounds}) gives unfortunately now a rather poor lower bound on the ground state energy $E_0 \geq -12-9{\bar n} -\ol{n^2}$, i.e. $E_0 \geq -12$.

\subsection{An anharmonic oscillator}

We next
 consider the Hamiltonian $H=(P^2+Q^2)/2+ \lambda Q^4/2 \geq 0$, $\lambda>0$,  to define the partition function. With the lower and upper symbols as given by Eqs.~(\ref{special}), (\ref{HQQ}), and (\ref{Q4}), we now find that
\b
\label{ZQ4upper}
Z(\beta) \leq \frac{1}{\beta\sqrt{2\pi}}e^{-\beta(k_2+\lambda a_4)/2}\int e^{-(x^2+\lambda(x^4/\beta+a_2x^2))/2}dx\; ,
\e
and
\b
\label{ZQ4lower}
Z(\beta) \geq \frac{1}{\beta\sqrt{2\pi}}e^{-\beta(\Delta(P)+\Delta(Q)+\lambda \ol{Q^4})/2}\int e^{-(x^2+\lambda(x^4/\beta+6x^2\ol{Q}^2))/2}dx \; .
\e
In the limit of large $\beta$, the lower bound on $Z(\beta)$ and the fact that $ H \geq 0$ then lead to $0\leq E_0 \leq (1+\lambda \ol{Q^4})/2$. With $\ss=\s|0\>\<0|$ one finds the upper bound $E_0 \leq (1+3\lambda/4)/2$ which, e.g.,  can be compared to the ``exact'' numerical value of $2E_0 =1.392351641530...$ for $\lambda=1$ \cite{banarjee_78}. We expect that this upper bound could be improved with a different choice of $\ss$.

A  consequence of the upper and lower bounds Eqs.~(\ref{ZQ4upper}) and (\ref{ZQ4lower}) now is that
for sufficiently small $\beta$ the upper and lower bounds converge to the well studied (see, e.g., Refs. \cite{borel} ) classical and asymptotic form
\b
\label{ZQ4exact}
Z(\beta)\hskip-1.3em&& =\frac{1}{\beta\sqrt{2\pi}}\int e^{-x^2/2- \lambda x^4/2\beta}dx \equiv Z_{cl}(\beta)/2\pi \no \\
&&=\frac{1}{2\lambda \sqrt{2\pi}}\sqrt{\frac{\lambda}{\beta}}e^{\beta/16\lambda}K_{1/4}(\beta/16\lambda)
\e
using Eq.~(\ref{Gibbs}) with $H_{cl}(p,q)= (p^2+q^2)/2+ \lambda q^4/2$. The expression in Eq.~(\ref{ZQ4exact}) involves all the energy states of the anharmonic oscillator in a highly non-trivial manner. In our case we are specifically interested in the limit $\beta \rightarrow 0$, i.e., $Z(\beta) \simeq \Gamma(1/4)(2\beta/\lambda)^{1/4}/2\beta\sqrt{2\pi}$.

\section{Comments}
For clarity, we have mainly focused on matrices $\ss=\ss(N)$ which meant that $\ss=\Sigma_{n=0}^\infty \s c_n\s |n\>\<n|$. More general matrices of course would involve expansions of the form
   \b \ss=\Sigma_{n,n'=0}^\infty\s c_{n,n'} |n\>\<n'| \e
   expressed in terms of a general matrix $\{c_{n,n'}\}$ that still ensures that $\ss$ has all the properties of a partition function. The use of such more general choices for $\ss$ will inevitably lead to expressions involving the matrix elements    \cite{Carruthers_1965}
       \b \label{eq:carruthers} && \hskip3cm\<n|\s U[p,q]\s|n'\> \no \\
&&\hskip-1cm= \sqrt{\frac{2^{n'}n!}{2^{n}n'!}}\exp [-\quarter(p^2+q^2)]\left( q+ip \right)^{n-n'}L_n^{n-n'}(\half(p^2+q^2))
\;, \e
 for $n \ge n'$ expressed in terms of the associated Laguerre polynomials Eq.~(\ref{Laguerre});
 instead, when $n<n'$, use $\<n|\s U[p,q]\s|n'\>= \<n'|\s U[-p,-q]\s|n\>^*$.
 The simple example where $\H=P^2+\omega^2\s Q^2$, $\omega\not=1$, shows that the optimal choice of $\ss$ is not always given by
   $|0\>\<0|$, where $(Q+iP)\s|0\>=0$, but in the present case by $\ss=|0;\omega\>\<0;\omega|$, where $(\omega\s Q+iP)\s|0;\omega\>=0$. This remark serves to confirm that the generalized representations have the possibility
   to make better bounds. It may be true that choices for $\ss$ of the form $|\psi\>\<\psi|$ (analogues of pure states) may be optimal, and
   that perhaps choosing $|\psi\>$ as the ground state of the Hamiltonian under examination may lead to
   optimal bounds. Those are interesting questions for the future.

         \section{Conclusion} We have developed new, classical, phase space bounds to deal with specialized
         (i.e., the partition function) questions that arise in quantum mechanics, and which, by their very nature, are technically easier to deal with than in their original form. It is quite likely that the generalized phase-space symbols we have introduced may have additional applications both in quantum mechanics
         and in time-frequency analysis.

\end{document}